\documentclass[12pt]{article}
\newcommand{\ignore}[1]{}
\usepackage[english]{babel}
\usepackage{graphicx,rotating,epsfig}
\usepackage{footmisc}

\newcommand{\TeV}{\ensuremath{\mathrm{Te\kern -0.1em V}}}
\newcommand{\GeV}{\ensuremath{\mathrm{Ge\kern -0.1em V}}}
\newcommand{\MeV}{\ensuremath{\mathrm{Me\kern -0.1em V}}}
\def\GeVc2{\ensuremath{\mathrm{ Ge\kern -0.1em V }\kern -0.2em /c^2 }}

\newcommand{\MW}{\ensuremath{M_{\mathrm{ W }}}}
\newcommand{\GW}{\ensuremath{\Gamma_{\mathrm{ W }}}}

\newcommand{\RunZ}{\hbox{Run-0}}
\newcommand{\RunI}{\hbox{Run-I}}
\newcommand{\RunII}{\hbox{Run-II}}

\begin{document}
\begin{center}
{\LARGE FERMI NATIONAL ACCELERATOR \\ \ \\ LABORATORY}
\end{center}

\begin{flushright}
       TEVEWWG/WZ 2009/01 \\
FERMILAB-TM-2439-E \\
       CDF Note 9859 \\
       D0 Note 5965 \\[1mm]
       10$^{\rm th}$ August 2009 \\
\end{flushright}

\vskip 1cm

\begin{center}
{\Large\bf Updated Combination of CDF and D0 Results \\
                  for the Mass of the $W$ Boson\\}

\vfill

{\Large
The Tevatron Electroweak Working Group\footnote{
The Tevatron Electroweak Working group can be contacted at tev-ewwg@fnal.gov.\\
 \hspace*{0.20in} More information is available at {\tt http://tevewwg.fnal.gov}.}\\
for the CDF and D0 Collaborations}

\vfill

{\bf Abstract}

\end{center}

{
We summarize and combine the results on the direct measurements of the mass of the $W$ boson
in data collected by the Tevatron experiments CDF and D0
at Fermilab. Results from CDF {\RunZ} (1988-1889) and {\RunI}
(1992-1995)  have been 
combined with D0 results from {\RunI}, the CDF 200 pb$^{-1}$ published results
 from the first period of {\RunII} (2001-2004) and the recent 1 fb$^{-1}$
result in the electron channel from D0 (2002-2006). The
results are corrected  for any inconsistencies  in   parton
distribution functions and assumptions about electroweak parameters used in the different analyses.  The resulting
Tevatron average for the mass of the $W$ boson
is $\MW = 80,420\pm 31~\MeV$.

\vfill



\section{Introduction}

The CDF and D0 experiments  at the Tevatron
proton-antiproton collider located at the Fermi National Accelerator
Laboratory  have made several direct measurements of the width, $\GW$, and  mass, $\MW$,
  of the $W$ boson.  These measurements use both the 
$e\nu$ and $\mu\nu$ decay modes of the $W$.

Measurements of $\MW$ have been reported by CDF from the data of
{\RunZ}~\cite{MW-CDF-RunZ}, {\RunI}~\cite{MW-CDF-RUN1A, MW-CDF-RUN1B}
and recently {\RunII}~\cite{MW-CDF-RUN2} and by
D0 from {\RunI}~\cite{MW-D0-I,MW-D0-I-rap,MW-D0-I-edge}. This document adds
a new D0 measurement from  {\RunII}~\cite{MW-D0-RUN2}.
There are no new measurements of the  width of the $W$ since   the previous average in July 2008\cite{TEV08}
and it will not be discussed further.

This note reports  the combination of the mass measurements,   and takes into account statistical and
systematic uncertainties as well as  correlations among
systematic uncertainties. It supercedes the  previous
summaries~\cite{MWGW-RunI-PRD,TEV08}.  The measurements are
combined using the analytic BLUE method~\cite{Lyons:1988,
Valassi:2003} which  is  mathematically equivalent to the methods used in  previous
combinations~\cite{MWGW-RunI-PRD,TEV08}, but, in addition,   yields
the decomposition of the uncertainty  on the average in terms of  
categories specified in the input measurements~\cite{Valassi:2003}.  These changes
are described in more detail in Ref. \cite{TEV08}.

As in the July 2008  analysis \cite{TEV08},  there are three significant changes relative to the pre-2008 averages:
\begin{itemize}
\item The individual $\MW$ measurement channels for CDF Run-0, Run-Ia
and Run-Ib  are now combined for each run period using the BLUE
method to achieve a consistent statistical treatment across all 
results. The \RunI\ D0 and  \RunII\ CDF measurements were already combined  
using the BLUE method.
\item  The central values of the  mass measurements made with very old PDF sets are corrected to
use the same parton distribution functions (PDFs)  from  CTEQ6M~\cite{CTEQ} with uncertainty estimates from the CTEQ6M, CTEQ6.1M~\cite{CTEQ61M} and  MRST2003~\cite{MRS} PDF sets\footnote{The CDF \RunZ\ and Run-Ia results were obtained from very old PDF sets 
(MRS-B~\cite{MRS-B} and \linebreak \nobreak{MRS-D$^\prime$~\cite{MRS-D}} respectively)
that did not utilize the $W$ charge asymmetry results and so 
provide somewhat offset predictions from the more modern PDF sets used
in the later analyses. The predictions based on the more modern
MRS~\cite{MRS} and CTEQ~\cite{CTEQ} sets used in Run-Ib and \RunII\ analyses 
have a variance smaller than the common PDF errors assumed in these analyses i.e.
$\approx 10$~MeV. We therefore only apply PDF corrections to the CDF
\RunZ\ and Run-Ia results since the shifts for these data are larger
than 10~MeV. 
We  retain the
PDF uncertainties of 60~MeV and 50~MeV quoted in the original
publications.  
We note that these corrections were also applied in the
\RunI\ CDF combination presented in~\cite{MW-CDF-RUN1B}.}.

The new D0 Run-II measurement uses CTEQ6.1M while the CDF Run II measurement used CTEQ6M. The difference in the mass extracted using these PDF sets was found to be less than $1 \pm 4$ MeV,
where 4 MeV is the statistical uncertainty on the estimated difference.  No correction is applied for this difference.

\item The mass values are  also corrected  to the
same assumed $W$ boson width value in order to achieve consistency across all results. The value of \GW\  quoted here corresponds to a definition 
based on a Breit-Wigner propagator in the ``running-width scheme", 
$1/{(M^2-\MW^2+iM^2\GW/\MW)}$, with a 
width parameter,   $\GW = 2093.2 \pm 2.2 $ MeV predicted by the standard model ~\cite{PR} using the 2008 world average $W$ boson mass of $80,398\pm 25 MeV$.
All measured masses  are corrected to this value using  $\Delta M_W  = -(0.15\pm0.05) \Delta\GW$ as in Ref.~\cite{TEV08}.
The $W$ boson mass uncertainty arising from an uncertainty in
the $W$ boson width is   now consistently treated across all measurements. 

\end{itemize}

The shifts due to these corrections are shown in Table  \ref{tab:MW-inputs09} below, which summarizes all of the inputs to the combination. \footnote{As described in \cite{TEV08}, the use of the BLUE method in internal combinations  caused the early  {\RunZ}, Run-Ia and Run-Ib CDF \MW\ values to change by 
$-3.5$~MeV, $-3.5$~MeV and $+0.1$~MeV respectively. These  corrections are also listed in
Table~\ref{tab:MW-inputs09}. When these new values are  combined using the BLUE method, the Run-0/I CDF combination is changed from $80,433 \pm\ 79$~MeV quoted
in~\cite{MW-CDF-RUN1B} and used in previous
combinations~\cite{MWGW-RunI-PRD}, to $80,436 \pm\ 81$~MeV.}

\begin{table}[!hbp]
\begin{center}
\begin{tabular}{|l |r |r |r |r |r |r |} \hline
 & CDF 0  & CDF Ia     & CDF Ib   & D0 I   & CDF II  & D0 II \\ \hline \hline
 $M_{W}$ published     & 79,910~~   & 80,410~~   & 80,470~~   & 80,483~~   & 80,413~~   & 80,400.7  \\ 
 Total uncertainty published &   390~~   &   180~~   &   89~~    &    84~~   &  47.9   &    43~~  \\ 

 $\Gamma_W$ used in publication      &  2,100~~   &  2,064~~   &  2,096~~   &  2,062~~   &  2,094~~   &  2,099.6 \\ 
\hline
\hline
{\bf Corrections}&&&&&&\\
\hline
$\Delta\Gamma_W$ correction applied &1.1& -4.4& 0.5& -4.7 &0.2& 1.0\\
PDF correction applied &20~~& -25~~& 0~~~& 0~~~& 0~~~& 0~~~\\
BLUE correction applied &-3.5& -3.5 &-0.1& 0~~~&0~~~&0~~~\\
\hline
Total correction     &   17.6  &  -32.9  &   0.4   &  -4.7   &   0.2   &   1.0   \\ \hline  \hline
\bf $M_{W}$ corrected     & 79,927.6 & 80,377.1 & 80,470.4 & 80,478.3 & 80,413.2 & 80,401.7 \\    
\hline \hline
{\bf Uncertainties}&&&&&&\\
 Total BLUE uncertainty     &  390.9  &  181.0  &   89.3  &  83.4   &  47.9   &   43.3  \\
 Uncorrelated uncertainty    &  386.1  &  172.8  &   87.9  &  82.1   &  44.7   &   41.3  \\   
\hline
 PDFs          &   60~~    &   50~~    &   15~~    &   8.1   &  12.6   &   10.4 \\
 Radiative corrections  &   10~~    &   20~~    &    5~~    &  12~~     &  11.6   &   7.5  \\
 $\Gamma_W$ (published)        &   0~~  &  20~~    &  0~~   &  10~~    &   0~~   &   0.5  \\ 

 $\Gamma_W$  (this analysis)        &   0.5   &  1.5    &  0.5    &  1.5    &   0.5   &   0.5  \\

 \hline \hline
\end{tabular}
\caption{Table 1 of the 2008 summary~\cite{TEV08}, updated with the   D0 result from  \RunII.  All entries are in MeV.
   \label{tab:MW-inputs09}}
\end{center}
\end{table}

\section{New data from D0 on $\MW$}

The  $W$ boson mass   determined by D0 in Run-II~\cite{MW-D0-RUN2}, using $W$
decays into electrons and neutrinos from 1 fb$^{-1}$ of data  at $\sqrt{s}=1960$ GeV, 
  derives from 3 observables: the transverse mass $M_T$, the electron transverse momentum $p^e_T$ and the transverse missing momentum, which   yield a combined $W$ boson mass of $80,401 \pm 21 (stat.) \pm 38 (syst.) $ MeV.  The individual   contributions to the uncertainty are
summarized in Table \ref{table:err}.


\begin{table}[!hbp]
\begin{center}
\begin{tabular}{|l |r|r|} \hline \hline
Source &Uncertainty in MeV& Correlation coefficient with \\
 && other experiments \\\hline
\hline
\bf Experimental uncertainties \rm&&\\
\hline
W Statistics &21.0& 0\\  
\hline
Electron energy calibration& 33.4&0\\  
Electron resolution model& 2.2&0\\
Electron energy offset & 5.2&0\\
Electron energy loss model& 4.0&0\\
Recoil model& 7.8&0\\
Electron efficiencies & 5.2&0\\
Backgrounds   & 3.2&0\\
\hline
\bf Production uncertainties\rm &&\\
\hline
PDFs& 10.4& 1.0\\
EWK radiative corrections & 7.5&1.0\\
Boson $p_T$  & 2.7 &0\\
$\Gamma_W$ & 0.5 &1.0\\ \hline
\end{tabular}
\caption{Contributions (in MeV) to the uncertainty for the  D0 Run-II $W$ boson mass result. \label{table:err}}
\end{center}
\end{table}

\section{Correlation of the D0 Run II result with other measurements}

The experimental systematic uncertainties on the new D0 measurement are dominated by 
the energy scale for electron candidates and are almost purely statistical, as they are mainly derived from the limited
sample of $Z^0$ decays.  All of the experimental uncertainties  are assumed to be uncorrelated
with previous measurements.

Three systematic uncertainties due to the production of  $W$ and $Z$ bosons are assumed to be fully correlated between all
Tevatron measurements, namely (1)
 the parton distribution functions (PDFs),
 (2)  the width of the $W$ boson ($\Gamma_W$) and
(3) the electroweak radiative corrections.

The D0 measurement also   includes an uncertainty in the boson $p_T$ distribution parameterization which is derived from a global fit to deep-inelastic scattering and hadron collider data~\cite{ref:Landry}.  In previous analyses, this source of uncertainty is treated differently, and it is therefore regarded as uncorrelated with
the earlier measurements.
 
Current estimates of the uncertainties due to radiative corrections include a significant statistical component.  The WGRAD/ZGRAD~\cite{ref:WGRAD} and PHOTOS~\cite{ref:PHOTOS} models are used in the different measurements and yield   results consistent within statistical uncertainties. We assume that    the  effects of radiative corrections are 100\% correlated between measurements because the models used are quite similar,  but we anticipate that both the uncertainties and the correlations have the potential to be reduced  in the future using better models and higher statistics in simulations.


\section{Combination of Tevatron $\MW$ measurements}
\label{cov}

The six measurements of $\MW$ to  be combined are given in
Table~\ref{tab:MW-inputs09}. The CDF \RunZ, Run-Ia and Run-Ib values correspond to averages of two measurements in different channels where
internal correlated systematic uncertainties  e.g. momentum scale, are
accounted for in the averaging. The \RunI\ D0 measurement combines
10   measurements using the BLUE method. The \RunII\ CDF
measurement combines a total of six individual measurements in the muon and electron decay channels.

Tables 1 and 2 of our previous summary~\cite{TEV08} are now extended with
the combined D0 \RunII\ result. The combined Tevatron result is calculated using  BLUE with input from
Table~\ref{tab:MW-inputs09} which includes the new D0 \RunII\ result.

The combined Tevatron result for the mass of the  $W$ is:
\begin{equation}
M_{W} = 80,420 \pm 31 \ \textrm{MeV}\,.
\end{equation}
The $\chi^{2}$ for the combined result of 2.69 for 5 degrees of freedom, corresponds to a
probability of 74.8\%.  Table \ref{contribution} shows the weight of
each measurement entering the combination.  

\begin{table}[!hbp]
\begin{center}
\begin{tabular}{|c |c |}  \hline 
           &    Relative Weights in \%  \\ \hline \hline
CDF 0      &   0.10  \\
CDF Ia     &   0.60  \\
CDF Ib     &   9.39  \\
D0 I     &  10.98  \\
CDF II     &  34.64  \\ 
D0 II    &  44.28  \\ \hline \hline
\end{tabular}
\caption{Relative weights of the contributions in \%. \label{contribution}}
\end{center}
\end{table}

The total uncertainty of 31 MeV on the Tevatron average is split up into an
uncorrelated uncertainty of 26.9 MeV, and  systematic uncertainties due to assumptions
about  the production   of  $W$ and $Z$ bosons of 11.6
(PDF), 9.3 MeV (radiative corrections), and 0.8 MeV ($\Gamma_W$).  The
global correlation matrix for the 6 measurements is shown in
Table~\ref{global}.

\begin{table}[!htp]
\begin{center}
\begin{tabular}{|c |l |l |l |l |l |l |} \hline
  & CDF 0  &CDF Ia    &CDF Ib   & D0 I   &CDF II  &D0 II \\
\hline  \hline
CDF 0 & 1.0  & 0.045 & 0.027 & 0.019 & 0.047  & 0.041 \\ \hline
CDF Ia  &    & 1.0   & 0.053 & 0.043 & 0.100  & 0.086 \\ \hline
CDF Ib  &    &       & 1.0   & 0.024 & 0.058 & 0.050 \\ \hline
D0 I  &    &       &       & 1.0   & 0.061 & 0.049 \\ \hline  
CDF II  &    &       &       &       & 1.0   & 0.106 \\ \hline
D0 II &    &       &       &       &       & 1.0  \\ \hline
\end{tabular}
\caption{Correlation coefficients between the different experiments using the method of Ref.~\cite{TEV08}. \label{global}}
\end{center}
\end{table}

\section{Conclusion}

The new   direct measurement of the mass of the  $W$ by the D0 experiment has been
combined with the previous CDF and D0 measurements.  The new
 Tevatron result for the $W$ boson mass is:
\begin{equation}
M_{W} = 80,420 \pm 31 \ \textrm{MeV}\,.
\end{equation}
For the first time the total uncertainty of  31 MeV from the Tevatron is
smaller than that of 33 MeV from    LEP II~\cite{LEPEWWG07}.

The combination of the new Tevatron result with the LEP II preliminary
result, assuming no correlations, yields the  
world average:
\begin{equation}  
M_{W} = 80,399 \pm 23 \ \textrm{MeV}\,.  
\end{equation}
Figure~\ref{fig:sum} shows an update of the Figure 1 of the TEVEWWG
note \cite{TEV08} displaying the new results.

\begin{figure}[hbt]
\centering \includegraphics [width=.90\textwidth] {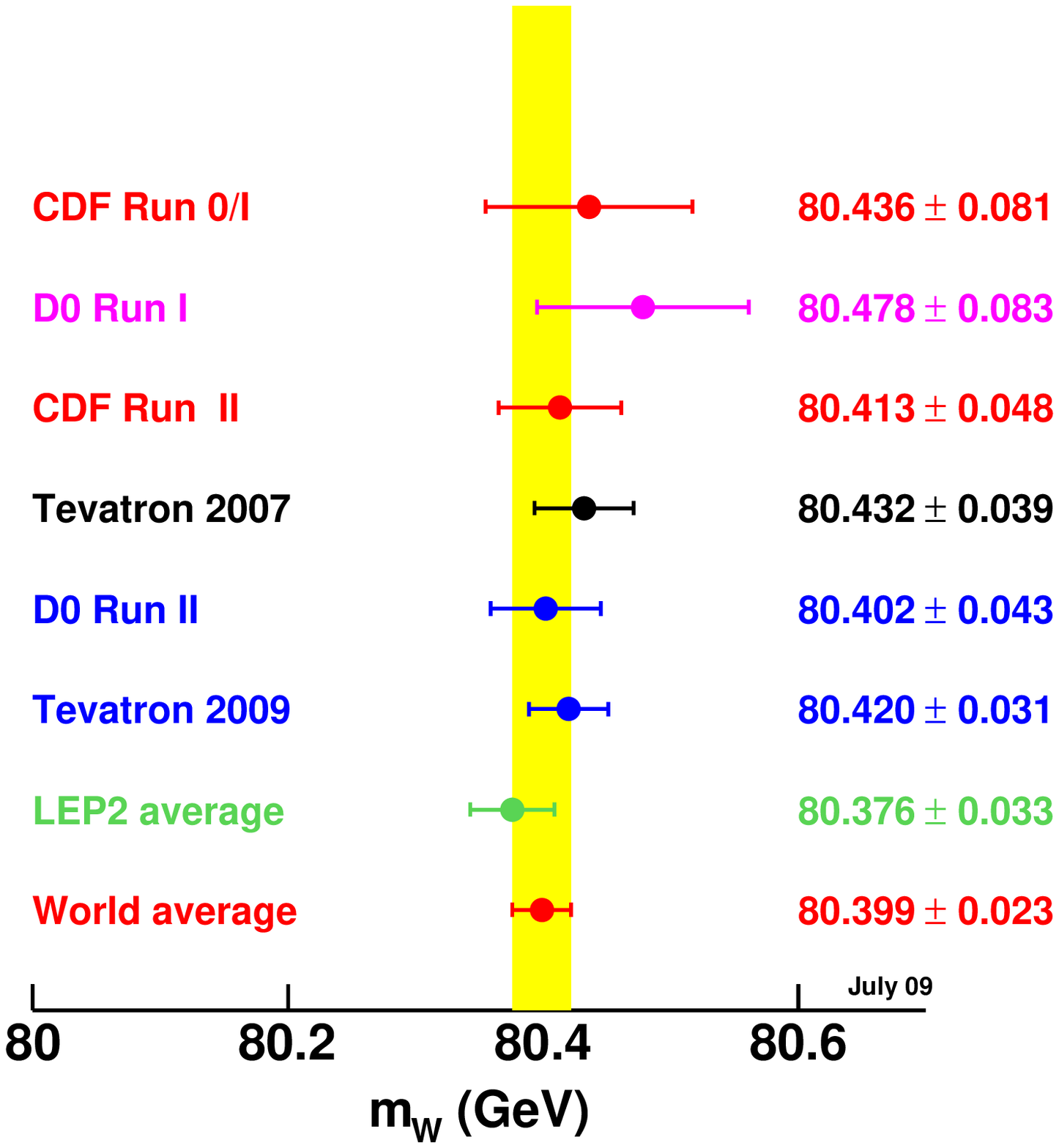}
\caption{
Summary of the measurements of the $W$ boson mass and their
average as of July 2009. The result from the Tevatron corresponds to  the values in this note
  (see Table
\ref{tab:MW-inputs09}) which include corrections to the same $W$ boson width and PDFs.  The LEP II result is from Ref.~\cite{LEPEWWG07}. An estimate of the world average of the Tevatron and LEP results assuming
no correlations between the Tevatron and LEP  is 
included.}
\label{fig:sum}
\end{figure}

\clearpage


\end{document}